\documentclass[12pt]{article}

\usepackage[margin=0.65in]{geometry}            
\usepackage{setspace}
\usepackage[modulo,mathlines,switch]{lineno}  
\usepackage{newtxtext,newtxmath}              
\usepackage[hidelinks]{hyperref}              
\usepackage{graphicx}                         
\usepackage[labelfont=bf]{caption}            
\usepackage[none]{hyphenat}
\usepackage{gensymb}
\usepackage{booktabs}
\usepackage{multirow} 
\usepackage{xcolor}
\usepackage{float}
\usepackage[inline]{enumitem}

\usepackage{caption}
\usepackage[
  backend=biber,
  style=numeric-comp,   
  sorting=none,         
  sortcites=true,       
  defernumbers          
]{biblatex}
\let\cite\supercite

\let\textcite\supercite

\defbibfilter{onlymain}{ segment=0 }
\defbibfilter{onlymethods}{ segment=1 and not segment=0 } 

\definecolor{lnred}{RGB}{220,0,0}  
\modulolinenumbers[1]                 
\newcommand{\wordrecognizermodel}{\emph{Word Recognizer}}
\newcommand{\sentencereadingmodel}{\emph{Sentence Reader}}
\newcommand{\textreadingmodel}{\emph{Text Reader}}

\begin{document}
\linenumbers        


\begin{center}
\textbf{\Large Hierarchical Resource Rationality Explains Human Reading Behavior}\\[8pt]

\textbf{Yunpeng Bai}$^{1,2}$,
\textbf{Xiaofu Jin}$^{2,3}$,
\textbf{Shengdong Zhao}$^{4}$,
\textbf{Antti Oulasvirta}$^{2,*}$\\[6pt]

$^{1}$National University of Singapore \;
$^{2}$Aalto University \;\\
$^{3}$Hong Kong University of Science and Technology (Guangzhou) \;
$^{4}$City University of Hong Kong\\[6pt]

*Corresponding author:\href{https://users.aalto.fi/~oulasvir/}{antti.oulasvirta@aalto.fi}
\end{center}


\noindent
\textbf{%
Reading is a pervasive and cognitively demanding activity that underpins modern human culture. It is a prime instance of a class of tasks where eye movements are coordinated for the purpose of comprehension.
Existing theories explain either eye movements or comprehension
during reading, but the critical link between the two remains unclear. 
Here, we propose resource-rational optimization as a unifying principle governing adaptive reading behavior. 
Eye movements are selected to maximize expected comprehension while minimizing cognitive and temporal costs, organized hierarchically across nested time scales: fixation decisions support word recognition; sentence-level integration guides skipping and regression; and text-level comprehension goals shape memory construction and rereading.
A computational implementation successfully replicates an unprecedented range of findings in human reading, from lexical effects to comprehension outcomes. 
Together, these results suggest that resource rationality provides a general mechanism for coordinating perception, memory, and action in knowledge-intensive human behaviors, offering a principled account of how complex cognitive skills adapt to limited resources.
}


\section*{}

\noindent  
Humans routinely engage in complex cognitive activities that require coordinating perception, memory, and action over time, under limited resources of attention, memory, and available time~\cite{lieder2020resource,simon1955behavioral,simon1956rational}. Explaining how such behavior remains adaptive when decisions are sequential and uncertain remains a central challenge for cognitive science~\cite{anderson2013adaptive,kahneman2011thinking}. While normative theories of bounded cognition characterize optimal behavior under constraints, how these principles operate in naturalistic, knowledge-intensive tasks is still poorly understood~\cite{simon1955behavioral,lieder2020resource}.

Reading exemplifies this class of resource-constrained, sequential behaviors: a ubiquitous and cognitively demanding activity in which eye movements and memory must be coordinated over time to support comprehension.~\cite{rayner1998eye}.
Beyond its practical importance, reading provides a naturalistic setting in which to study how humans allocate attention and memory over time to support comprehension under resource constraints~\cite{kintsch1978toward,rayner1998eye}.
Computational approaches to explain reading have proceeded along two complementary paths.
The first describes how cognition guides eye movements~\cite{reichle2003ez,salvucci2001integrated,legge1997mr,norris2006bayesian,bolliger2023scandl,Bolliger2025ScanDL2}, and the second focuses on how memory and comprehension are constructed~\cite{kintsch1978toward,kintsch1978toward,van1999landscape,tapiero2007situation}. Within the first path, word recognition is assumed to proceed serially with shifts of attention~\cite{reichle2003ez}, with lexical processing stages directly triggering eye movements, 
as being distributed across several words at once, with eye movements resulting from a stochastic competition between their activations~\cite{salvucci2001integrated}, 
or as an uncertainty-reduction problem: eye movements are chosen to minimize entropy over word identity~\cite{legge1997mr,norris2006bayesian}. 
These approaches focus on capturing local lexical control, leaving unclear how eye movements are guided by evolving comprehension and memory demands over time.
Approaches using a machine learning technique called conditional generation~\cite{bolliger2023scandl,Bolliger2025ScanDL2} treat fixations as a mapping from sentence's word orders and semantic meanings. While this approach shows high predictive accuracy, it lacks explicit cognitive assumptions about memory and control. 
The second path comprises theories of comprehension in reading~\cite{yeari2016computational,kintsch1978toward,van1999landscape,tapiero2007situation}, which center on the construction-integration process. It is assumed that a schema guides working memory in building a coherent propositional text base and reducing it into a gist. Although these theories capture the formation of meaning, they do not explain how this, in turn, influences eye-movement control. 
To sum up, no existing work explains how comprehension---beyond lexical activation and word meaning--and eye-movement control interact. A unified account of this process is missing. 

In this work, we provide strong evidence for a unified computational basis underpinning reading as a joint optimization of eye movement and comprehension. 
We propose that resource-rational control~\cite{lieder2020resource,gershman2015computational,howes2009rational} provides a general theoretical account of how humans coordinate perception, memory, and action to maximize long-term utility under limited resources during reading, and we instantiate this claim in a computational model.
%
%
Fundamentally, reading is a process of sequential information sampling: each fixation contributes evidence that updates memory and supports comprehension, while future eye movements are chosen in light of these evolving internal states. Moreover, evidence from eye-tracking and neuroscience further shows that cognition is organized hierarchically across multiple time scales~\cite{simony2016dynamic,wang2025fast,pallier2011cortical}. Accordingly, the principle of resource rationality can be extended to derive hierarchical action representations that balance behavioral efficiency with flexibility~\cite{bacon2017option,lieder2020resource}. These principles naturally link high-level cognition with low-level eye movements: fixations emerge as consequences of boundedly optimal actions selected to maximize comprehension goals. Sequential sampling under hierarchical control ensures that low-level eye movements strategically accumulate information for memory step by step, providing the foundation on which higher-level comprehension processes are built.

\begin{figure*}[th]
    \centering 
    \includegraphics[width=1.0\textwidth]{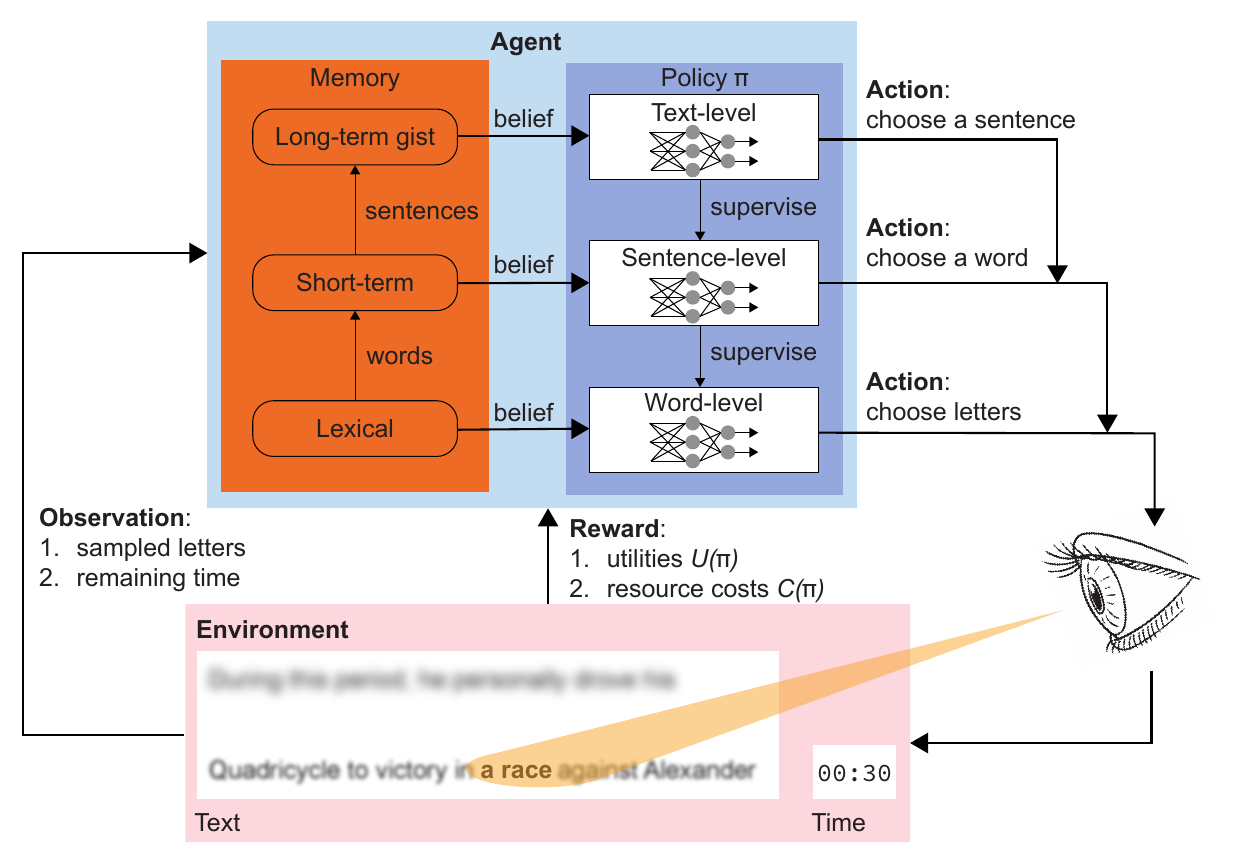} 
    \caption{
    \textbf{A resource-rational mechanism for reading.} The reader is simulated as a resource-rational~\cite{lieder2020resource,gershman2015computational} agent that maximizes text comprehension (goal) by optimizing fixations on a text display (external environment), under constraints of limited resources such as memory, visual perception, and time.
    The agent is uncertain about the true states because its observations are partial. To handle such uncertainties, it maintains probabilistic beliefs over the word, sentence, and text states through three memory stores -- Lexical, Short-term, and Long-term.
    Decisions are made hierarchically, high-level text comprehension guide sentence-level readings, which in turns guides word recognition and eye-movement control.
    }
    \label{fig:model overview} 
\end{figure*}

We develop a computational model grounded in these principles to simulate a resource-rational reader. The reader is formalized as an agent that optimizes comprehension through a sequence of decisions, trading off expected accumulative gains in understanding against the costs of eye-movement effort and time under limited visual, memory, and temporal resources (Fig.~\ref{fig:model overview}). It features a hierarchical control architecture with three interacting levels: word-, sentence-, and text-levels, each represented as a controller that contributes to adaptive reading behavior (Fig.~\ref{fig:hierarchical pomdp}). 

Our model is the first, to our knowledge, to jointly capture how comprehension guides fixations and how fixations support comprehension within a single computational principle.
It accurately reproduces a broad range of known empirical effects of eye movements and comprehension in reading from the principle of resource rationality~\cite{lieder2020resource,gershman2015computational}. At the word level (Fig.~\ref{fig:simulation gallery}a), it captures established findings~\cite{kliegl2004length,rayner1998eye,rayner2011eye,staub2015effect}: longer words prolong fixations, while frequency and predictability shorten them, reflecting eye movements are rational controls for reducing lexical identity uncertainty in memory. At the sentence level (Fig.~\ref{fig:simulation gallery}b), it predicts skips and regressions as rational trade-offs between efficiency and integration accuracy in the short-term memory. At the text level (Fig.~\ref{fig:simulation gallery}c), it explains how prior knowledge~\cite{mcnamara1996good,mcnamara2001reading} and coherence~\cite{mcnamara2001reading,mcnamara1996good,freebody1983effects} improve comprehension and trigger targeted regressions to repair failures in text-level comprehension~\cite{vasishth2013eye}. We collected a new dataset of English reading under time pressure and compared our model’s simulations against it. The model successfully captured human adaptations in skipping, regressions, and comprehension decline~\cite{duggan2011skim,vibert2025impact} (Fig.~\ref{fig:simulation gallery}d), revealing that readers strategically adjust eye-movement patterns to maximize comprehension under limited time and memory capacities.
Crucially, these behaviors are not prescribed by fixed heuristics or task-specific rules, but instead emerge from optimizing comprehension under hierarchical resource constraints, highlighting the explanatory necessity of a resource-rational account of reading.

\subsection*{Computational simulation of reading}\label{subsec:principles}
\begin{figure*}[p]
    \centering 
    \includegraphics[width=0.85\textwidth]{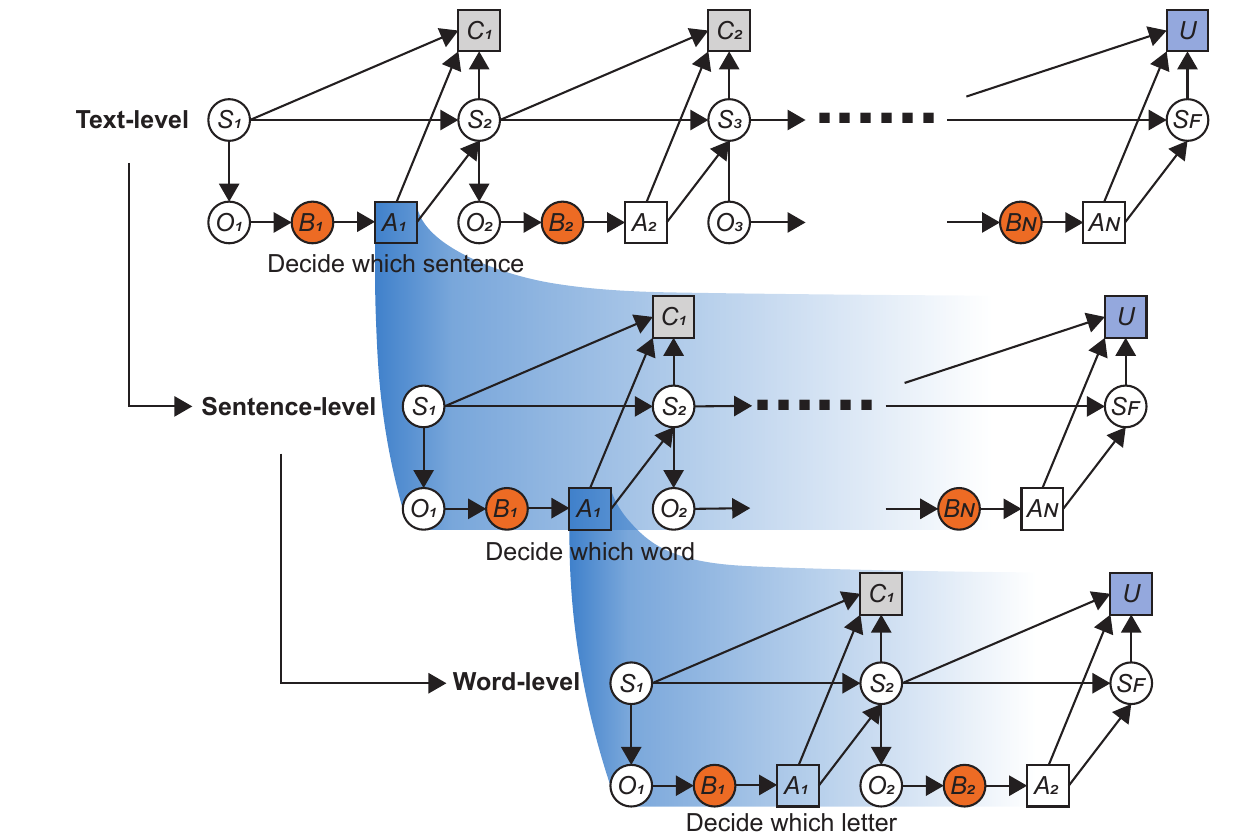} 
    \caption{
    \textbf{Hierarchical resource-rational control of reading formalized as POMDPs.}
    Eye-movement control is organized into three nested levels: Text, Sentence, and Word, each operating on its own temporal scale. 
    Higher-level controllers set coarse reading targets, and lower-level controllers execute them at finer scales. For instance, the text-level controller chooses the sentence, the sentence-level chooses the word within that sentence, and the word-level recognize that word by deciding how to sample the letters.
    At each level, the process is formalized as a POMDP. The state $S_N$ describes the reading dynamics; observations $O_N$ reflect noisy visual input sampled from the environment (text stimulus) due to limited visual attention, forming partial observability from $S_N$. Because the true state $S_N$ is not directly observable to the memory, the agent maintains a belief state $B_N$ based on $O_N$ in the memory to handle uncertainties: encoding, for example, coherence appraisals across sentences, word predictions, or word activation distributions. Actions $A_N$ are selected based on $B_N$ to advance reading, which in turn updates the state $S_N$ and produce new observations $O_N$ for the next step. $S_F$ represents the final step at given level.
    Decision-makings are resource-rational: at each level, the agent learns to select the action that maximizes expected reward, defined as comprehension utility $U$ and eye-movement costs $C_N$. 
    }
    \label{fig:hierarchical pomdp} 
\end{figure*}

We formalize resource-rational control hierarchically to capture how human readers naturally decompose the complex task of understanding text into nested subtasks during reading. The hierarchy consists of three levels operating from low to high (Fig.~\ref{fig:hierarchical pomdp}). (1) Word level: the controller decides which letters to fixate for word recognition, formalized as reducing uncertainty in word identity within the lexical memory. (2) Sentence level: the controller selects which word to read next to support sentence comprehension, balancing reading speed against information gain and comprehension accuracy. This is reflected in the trade-off between skips and regressions, primarily involving short-term memory processes. (3) Text level: the controller determines which sentence to read or reread to achieve coherent understanding across the text, coordinating long-term memory construction and coherence monitoring. The three controllers operate at different temporal and informational scales but interact continuously. For example, the sentence-level controller instructs the lexical controller which word to process next. The lexical controller then directs visual attention to that location, deploying fixations to sample letters and identify the word. The recognized word is passed back as perceptual evidence to the sentence-level controller, which updates its representation of sentence meaning and, in turn, contributes to text-level comprehension governed by the higher controller.

Each controller's control problem is represented as a Partial Observable Markov Decision Process (POMDP)~\cite{sutton1998reinforcement} (Fig.~\ref{fig:hierarchical pomdp}), a mathematical framework for modeling sequential decision-making under uncertainty. We adopt this formulation to describe reading's resource-rational control because reading is inherently sequential: at each step, the agent decides where to look next based on its current internal state, including what has been seen, remembered, and understood so far, and its progress toward comprehension. The POMDP also captures the uncertainty and partial observability intrinsic to reading. Readers cannot fully observe the text at once; letters and words outside fixation are only partially visible, so decisions must be made under uncertainty using imperfect sensory and memory information. Within this formalism, the belief state compactly summarizes ongoing comprehension and guides the next eye-movement action. Each controller is trained independently at its own level using deep reinforcement learning~\cite{schulman2017proximal}, which learns boundedly optimal policies that maximize comprehension utility while minimizing the costs of eye-movement effort and time. After training, the three controllers are integrated to function as a unified model capable of simulating natural reading behaviors.

\section*{Results}\label{sec:results}

We compare our model's simulation results against a set of established empirical effects at the word~\cite{rayner1998eye,kliegl2004length}, sentence~\cite{rayner1998eye,kliegl2004length,staub2007eye,vasishth2013eye}, and text levels~\cite{mcnamara1996good,kintsch1978toward}; as well as reading under time constraints~\cite{vibert2025impact}. These well-established effects have long been used to characterize the adaptive dynamics of human reading, reflecting how eye movements and linguistic processing interact under resource constraints. Successfully reproducing them provides a rigorous test of whether the model captures the hierarchical resource-rational mechanisms in reading.

\begin{figure*}[p]
    \centering 
    \includegraphics[width=\textwidth]{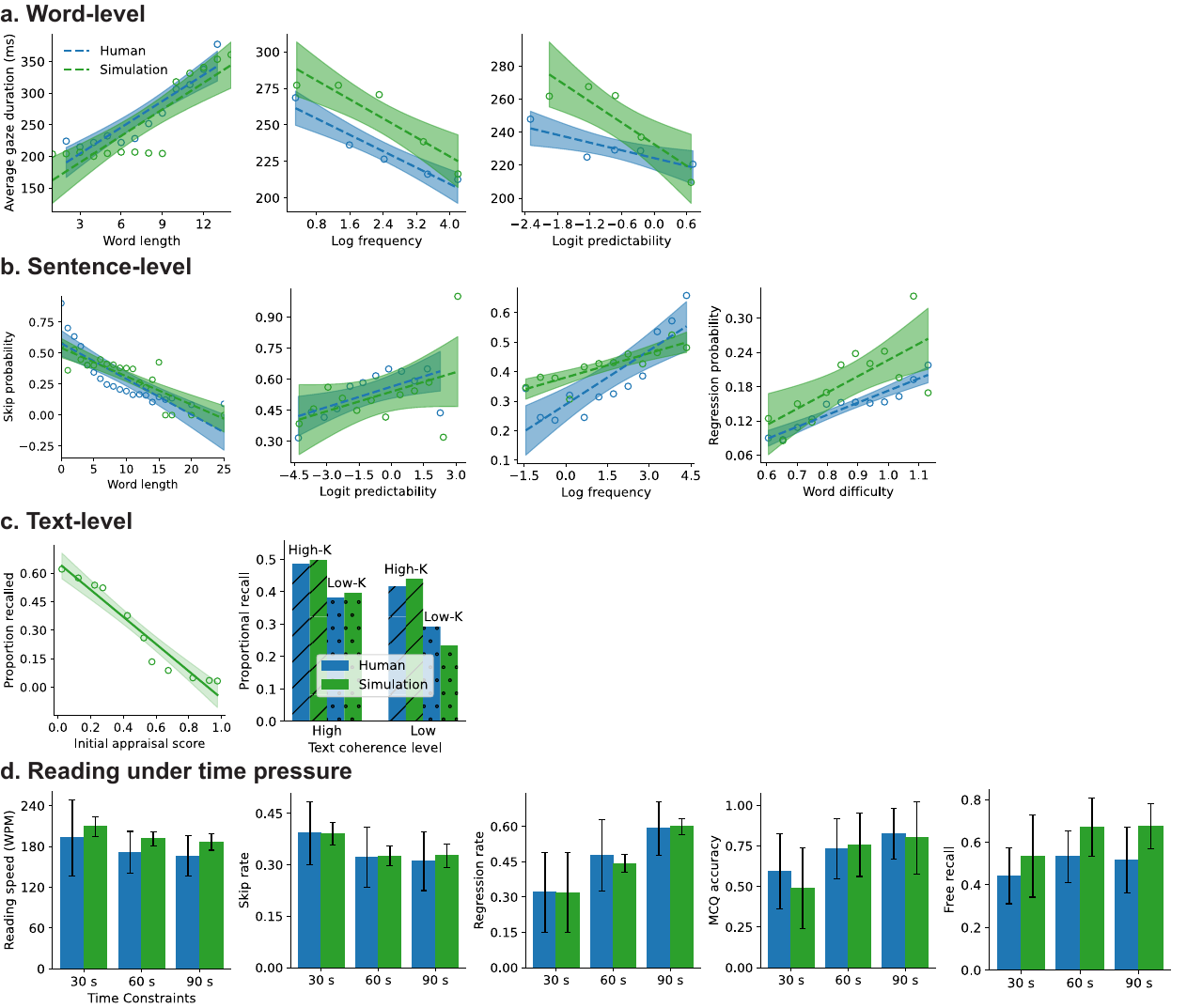} 
    \caption{
    \textbf{Model reproduces key empirical effects in eye movements and comprehension.}
    \textbf{a. Word-level.} Both humans and the simulation show lexical influences on gaze duration: shorter, more frequent, and more predictable words receive shorter gaze duration.  
    \textbf{b. Sentence-level.} Skips and regressions. Short, frequent, and predictable words are skipped more often, whereas lexically or contextually difficult words elicit more regressions.  
    \textbf{c. Text-level.} Comprehension-driven control. The simulation reproduces targeted regressions to poorly understood passages (low initial appraisal) and captures how prior knowledge and text coherence improve recall.  
    \textbf{d. Reading under time pressure.} Adaptive strategy shifts. Increasing time pressure leads to faster reading with more skips and fewer regressions, accompanied by reduced multiple-choice accuracy and free recall.  
    Human data are shown in blue and model predictions in green; dashed lines indicate linear fits with 95\% confidence intervals, and bars show means with standard deviations.  
    Together, these results demonstrate reading as a resource-rational process: readers adapt their eye movements to linguistic features and available time, producing corresponding adaptations in comprehension. These adaptations reflect a systematic trade-off between accuracy and effort to maximize overall utility across word-, sentence-, and text-level processing.
    }
    \label{fig:simulation gallery} 
\end{figure*}

\subsection*{Deciding when and where to fixate in a word}\label{sec2_subsec1}

Word-level eye-movement behavior reflects how readers manage uncertainty during lexical identification under limited visual, attentional, and memory resources.
Empirical findings show that gaze duration (GD), defined as the sum of all fixation durations on a word during its first pass, varies systematically with word length, frequency, and contextual predictability~\cite{rayner1998eye,kliegl2004length}. Our simulations match human effects (Fig.~\ref{fig:simulation gallery}a): GD increases with length (Human: $\beta=13.85$, $R^2=.85$; Simulation: $\beta=14.00$, $R^2=.73$), decreases with log frequency (Human: $\beta=-14.01$, $R^2=.92$; Simulation: $\beta=-16.27$, $R^2=.85$), and decreases with logit predictability (Human: $\beta=-7.83$, $R^2=.71$; Simulation: $\beta=-21.66$, $R^2=.79$), indicating that the model captures both direction and approximate magnitude of such lexical features' effects.

\begin{figure*}[th]
    \captionsetup{labelfont=bf,name=Extended Data Figure}
    \centering 
    \includegraphics[width=\textwidth]{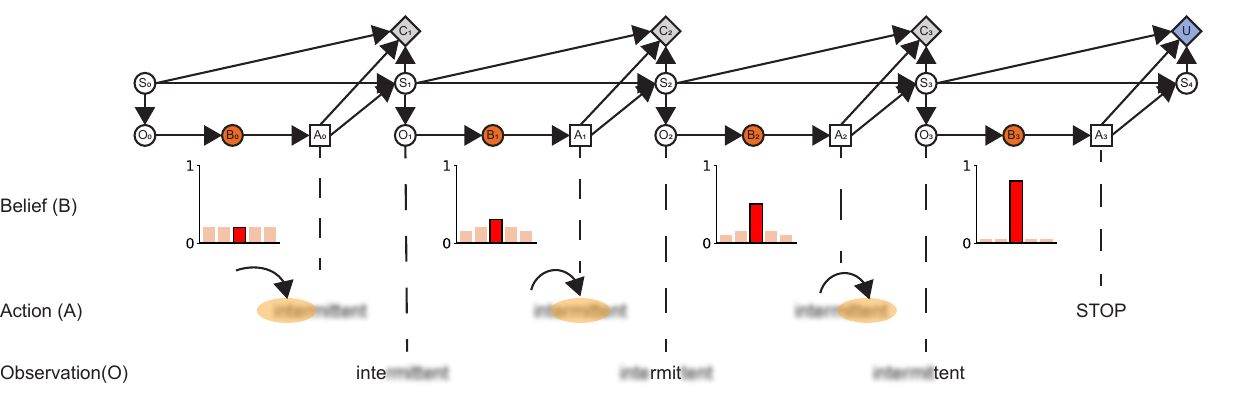} 
    \caption{
    \textbf{Illustration of word-recognition dynamics in the resource-rational POMDP model.}
    At each fixation, the agent receives a noisy visual observation $O_t$ (letters within the current foveal region), which updates a belief distribution over candidate words in lexical memory $B_t$ (orange nodes). The bar plots show these evolving belief states: as more evidence is sampled, the probability of the correct word (red bar) increases and becomes discriminable from candidates. Based on the current belief, the agent selects an eye-movement action $A_t$, either continuing to sample additional letters (and where to fixate) or terminating recognition. Frequent or predictable words require fewer fixations because their beliefs converge more quickly. Together, word recognition emerges as a boundedly optimal process of reducing uncertainty while balancing eye-movement costs $C_t$ and recognition utility $U$.    
    }
    \label{fig:extended data word recognition} 
\end{figure*}

We model word recognition as a resource-rational control problem framed as a POMDP (Extended Data Fig.~\ref{fig:extended data word recognition}). Because visual attention is limited, human readers often require multiple fixations on different parts of a word to identify it accurately. resource rationality explains this as uncertainty reduction: for both human and agent reader, partial visual input of letters from each fixation activates several candidate words in lexical memory, creating uncertainty about the word’s identity. To resolve this uncertainty efficiently, the reader must select fixation locations that maximize information gain. We implement this mechanism in the reading agent using Bayesian inference, a natural computational framework for human-like word recognition~\cite{norris2006bayesian}. The agent maintains a belief distribution over candidate words, updated after each fixation. Lexical frequency and contextual predictability serve as prior probabilities, while sampled letters provide likelihood evidence; together, they drive belief updates during inference. Fixation duration is modeled as proportional to uncertainty reduction: when multiple candidates remain active, fixations lengthen to resolve competition~\cite{norris2006bayesian}. At each step, the agent chooses actions (fixation locations and termination decisions) that balance recognition accuracy against the effort and time cost of eye movements.

This model explains classic word-level eye-movement effects as consequences of rational uncertainty reduction during lexical identification. Longer words yield greater lexical uncertainty, requiring more and longer fixations for identification. In contrast, frequent or predictable words begin with sharper priors which lead to more peaked belief distributions, and thus reach confident recognition more quickly, resulting in shorter and fewer fixations. Word recognition emerges as a boundedly optimal process of uncertainty minimization, in which fixation timing and location reflect rational trade-offs between information gain, word recognition accuracy, and eye-movement costs.

\subsection*{Deciding where to fixate in a sentence}

At the sentence level, readers must decide not only how long to fixate on words, but also whether to skip ahead or regress to earlier text in order to maintain coherent comprehension.
Lexical properties systematically shape eye-movement control at sentence-level: shorter, more frequent, and more predictable words are skipped more; difficult words elicit regressions~\cite{rayner1998eye,kliegl2004length,staub2007eye,vasishth2013eye}. 
We use the Task 2 of the ZuCo 1.0 dataset, which provides detailed eye-movement records during natural sentence reading~\cite{hollenstein2018zuco}. We derived two complementary measures to capture these effects: skipping probability, defined as the proportion of trials in which a word received no first-pass fixation, and regression probability, defined as the proportion of trials in which a word was refixated after the gaze had already moved past it. 
Our model reproduces these effects (Fig.~\ref{fig:simulation gallery}b). Skipping decreases with word length (Human: $\beta=-0.03$, $R^2=.70$; Simulation: $\beta=-0.02$, $R^2=.73$), increases with log frequency (Human: $\beta=0.06$, $R^2=.70$; Simulation: $\beta=0.02$, $R^2=.74$) and with logit predictability (Human: $\beta=0.03$, $R^2=.41$; Simulation: $\beta=0.03$, $R^2=.20$). Regression rates rise with word difficulty (Human: $\beta=0.21$, $R^2=.90$; Simulation: $\beta=0.29$, $R^2=.51$). 
Together, these results show that the model captures both the direction and approximate strength of lexical influences on skips and regression behavior during sentence reading.

\begin{figure*}[th]
    \captionsetup{labelfont=bf,name=Extended Data Figure}
    \centering 
    \includegraphics[width=\textwidth]{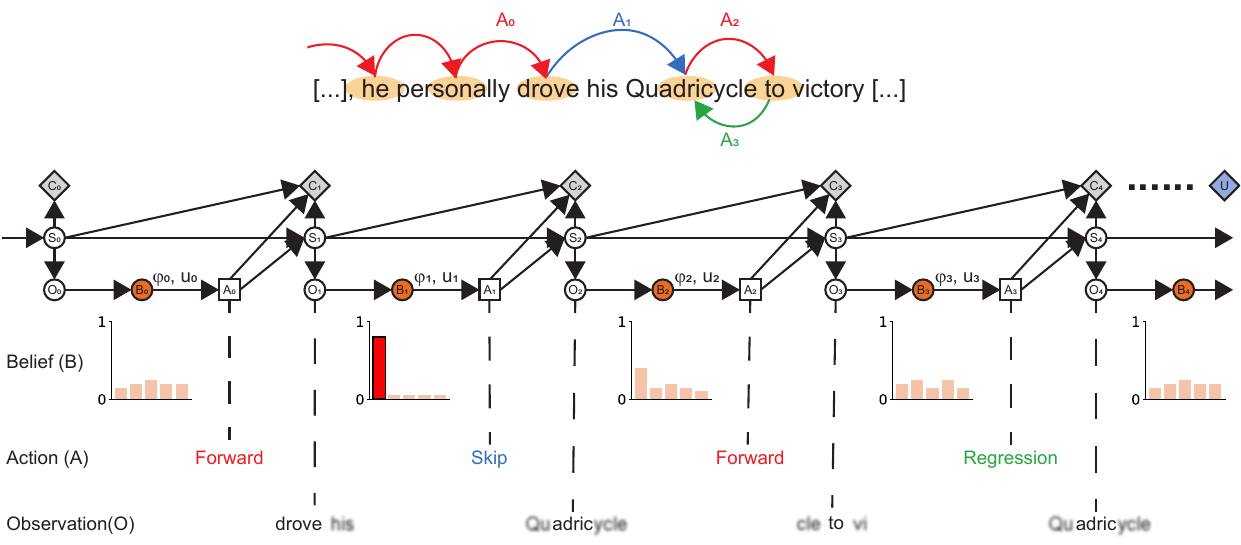} 
    \caption{
    \textbf{Illustration of sentence-level reading dynamics in the resource-rational POMDP model.}
    The agent reads sentences through a combination of forward movements, skipping, regressions, and termination decisions. When reading the given sentence, it minimizes eye-movement effort and time cost $C_t$ while maximizing sentence-level comprehension utility $U$, defined as $u_t$, which is the geometric mean of all word appraisals $\varphi_t$. Words that are highly predictable are skipped to save time, whereas difficult or incoherent words trigger regressions to restore comprehension. Visual input $O_t$, sampled by word recognition agent, includes both foveal and parafoveal information: letters at fixation are seen clearly, while adjacent words are perceived with lower acuity. This parafoveal preview, combined with contextual cues in short-term memory, allows the agent to predict upcoming words and maintain a belief distribution over possible candidates (shown as bar charts; higher confidence indicated by red bars). Based on these beliefs, the agent strategically decides whether to skip or make a forward fixation. Regression decisions, in contrast, are guided by the estimated comprehension of the text read so far. When comprehension drops, the agent evaluates whether the expected gain from rereading outweighs the additional time and effort costs, and will regress if doing so improves overall understanding. Together, these processes demonstrate how sentence reading emerges as a boundedly optimal control of visual sampling and memory integration under cognitive and temporal constraints.
    }
    \label{fig:extended data sentence reading} 
\end{figure*}

Sentence-level eye-movement control can be understood as strategic action selection under uncertainty, where readers balance comprehension goals against the costs of eye movements and time. 
We formalize eye-movement in sentence reading as resource-rational control under partial observability (Extended Data Fig.~\ref{fig:extended data sentence reading}). Rather than a linear or reflexive sequence~\cite{bicknell2010rational,reichle2003ez}, sentence reading is modeled as a strategic control problem in which the reader chooses among actions: moving forward, skipping, regressing, or terminating to optimize expected comprehension under visual attention and memory constraints. 
In this formulation, the agent’s decisions are guided by two internal beliefs. A comprehension belief tracks how well previously read words have been integrated into a coherent representation, while a prediction belief estimates the likelihood of upcoming words based on context and parafoveal information. When the comprehension belief is low, which quantifies poor semantic integration among read words (see Methods), the agent regresses to trade additional eye-movement and time costs for improved comprehension. This mechanism explains why difficult words, which disrupt integration, trigger more regressions. Conversely, when the prediction belief for upcoming words is high, which indicates strong confidence based on short-term memory context and parafoveal preview sampled letters, the agent skips these words to save time with minimal expected information loss. This accounts for greater skipping of short, frequent, or predictable words, which naturally generate higher prediction beliefs.
Crucially, unlike descriptive models that hard-code eye-movement functions~\cite{reichle2003ez,salvucci2001integrated}, where rational analyses of active skips are based on preset threshold~\cite{duan2020rational}, or black-box sequence generators to approximate patterns without interpretability~\cite{bolliger2023scandl,Bolliger2025ScanDL2}, our policy for deciding where to fixate in sentence reading emerges naturally from agent's active control. The model maximizes comprehension utility while minimizing the effort and time cost of eye movements under limited visual attention and memory resources. This resource-rational formulation provides a mechanistic explanation for how sentence-level reading behaviors (especially skips and regressions) arise as adaptive trade-offs between reading efficiency and comprehension accuracy.

\subsection*{Text comprehension and deciding where to read in text}

At the level of extended text, readers must decide not only how to process incoming information, but also when to revisit earlier material in order to maintain a coherent mental representation.
Prior work has found that both text coherence (the explicit connectedness of ideas within and across sentences) and prior knowledge (readers’ existing understanding of the text domain) strongly shape comprehension and recall~\cite{mcnamara1996good,kintsch1978toward}. Using McNamara et al.’s dataset~\cite{mcnamara1996good}, we replicated these effects and matched their magnitudes (Fig.~\ref{fig:simulation gallery}c): participants recalled a greater proportion of propositions when they possessed higher prior knowledge (Human: $M=0.45\pm0.03$; Model: $M=0.47\pm0.03$) compared to lower prior knowledge (Human: $M=0.34\pm0.05$; Model: $M=0.31\pm0.08$). Likewise, coherent texts -- those with stronger local and global connections—were remembered better (Human: $M=0.43\pm0.05$; Model: $M=0.45\pm0.05$) than low-coherence texts (Human: $M=0.35\pm0.06$; Model: $M=0.33\pm0.10$). Beyond reproducing mean recall scores, our model predicts an adaptive regression mechanism for optimizing comprehension: after an initial forward pass, the agent selectively revisits sentences with low inferred comprehension, strengthening integration and repairing ambiguities. This targeted rereading parallels human patterns of selective regressions rather than indiscriminate backtracking~\cite{vasishth2013eye}.

\begin{figure*}[th]
    \captionsetup{labelfont=bf,name=Extended Data Figure}
    \centering 
    \includegraphics[width=\textwidth]{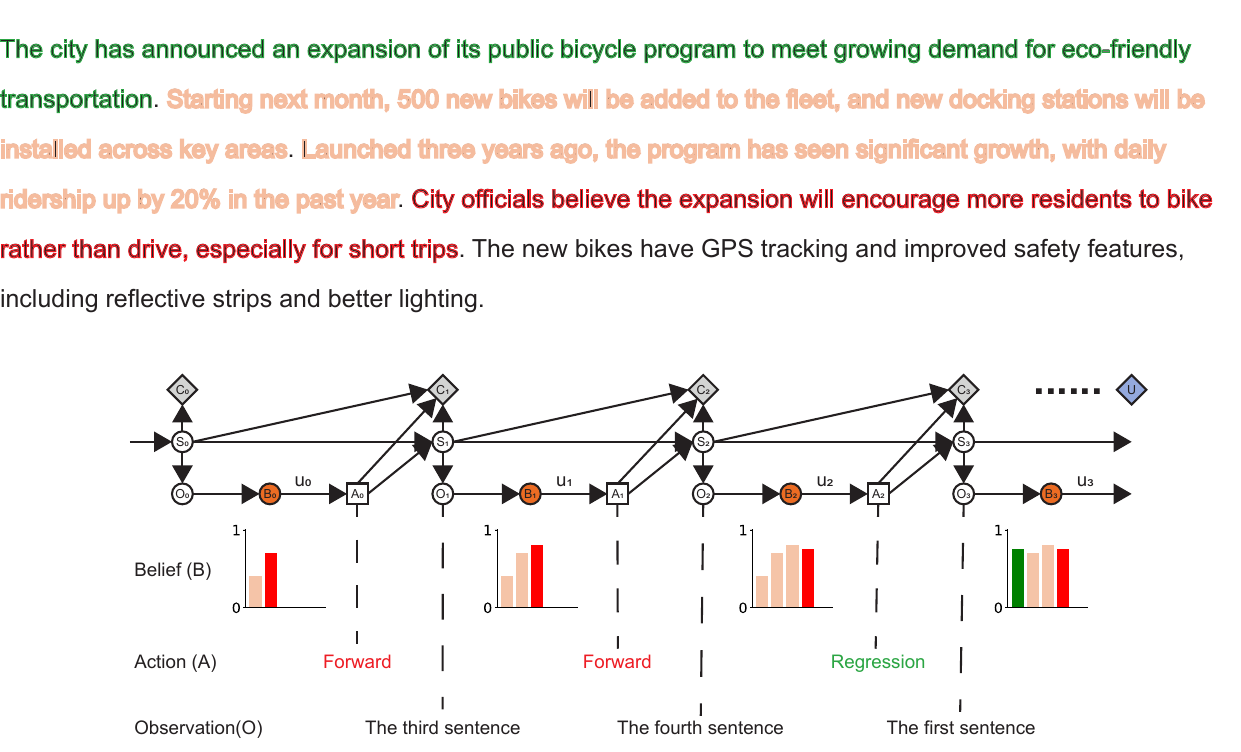} 
    \caption{
    \textbf{Illustration of text-level reading dynamics in the resource-rational POMDP model.}
    When reading multi-sentence text, the agent decides whether to proceed to the next sentence (red), regress to an earlier sentence (green), or terminate reading. These decisions are governed by two interacting components. (1) Sentence appraisals $B_t$ quantify how well each sentence integrates into the evolving schema-guided long-term memory; bar charts show the belief distribution over appraisals, with higher values indicating stronger contributions to coherence. (2) Text-level comprehension $u_t$ is computed as the geometric mean of all processed sentence appraisals. When comprehension of a newly read sentence falls below expectation, the agent strategically regresses to reread earlier sentences. Such revisits trigger additional processing cycles that strengthen relevant propositions in memory, raising their appraisal scores and improving global coherence. These dynamics illustrate how reading decisions arise from the continuous optimization of overall comprehension utility $U$ against the time and effort costs of eye movements $C_t$, it also demonstrates the mutual effect between eye-movement control and memory construction in text comprehension.
    }
    \label{fig:extended data text comprehension} 
\end{figure*}

At the text level, comprehension is governed by resource-rational control over an evolving, partially observable text state (Extended Data Fig.~\ref{fig:extended data text comprehension}).
Because the agent cannot directly observe the full underlying state of the text, it must rely on partial and memory-limited observations, derived from noisy letter sampling, capacity-constrained sentence integration, and compressed long-term representations. Thus, the agent must rely on a belief state that tracks the reader’s evolving interpretation of the text. 
At each step, the agent strategically chooses from the action space of continue reading, regress to an earlier sentence, or terminate, to optimize comprehension. Memory construction is modeled to parallel human processes inspired by Kintsch et al.'s framework~\cite{kintsch1978toward}: readers build coherence across sentences by coordinating short-term and long-term memory, but because working memory is capacity-limited, only a subset of propositions can be retained. Readers therefore prioritize propositions that are coherent or important while activating relevant knowledge from long-term memory to support integration. Propositions that persist across cycles are gradually consolidated into compressed ``gist” representations, forming the backbone of text-level understanding. The effect of prior knowledge is captured through the model’s activation threshold, which determines how relevant a proposition must be to be integrated into long-term memory. Higher thresholds indicate greater difficulty of activation, corresponding to readers with lower prior knowledge. In contrast to prior work, the effect of text coherence does not require a threshold parameter: when the agent parses texts of varying coherence, the computed proposition-to-proposition relevance naturally differs: coherent texts yield higher internal consistency, whereas incoherent texts produce lower relevance scores. Thus, coherence effects emerge intrinsically from the text structure itself. 

These memory dynamics directly shape eye-movement behavior. When integration difficulty or ambiguity arises, readers deploy targeted regressions as compensatory control~\cite{vasishth2013eye,staub2007eye}, strategically revisiting earlier text to restore coherence. In our model, such regressions are not reactive noise but rational investments, where costly actions taken only when the expected gain in comprehension outweighs the time and effort cost. This mechanism naturally explains why rereading increases under low prior knowledge or low coherence: when memory integration falters, the optimal policy favors revisiting earlier material to recover coherence. Unlike comprehension-only frameworks~\cite{van1999landscape,kintsch1978toward,tapiero2007situation}, which assume input is already encoded, our account links gaze allocation directly to memory integration, providing a principled computational explanation of how attention control and text-level comprehension are bidirectionally coordinated under resource constraints.

\subsection*{Speed-accuracy trade-off when reading under time pressure}

\begin{figure*}[p]
    \centering 
    \includegraphics[width=0.90\textwidth]{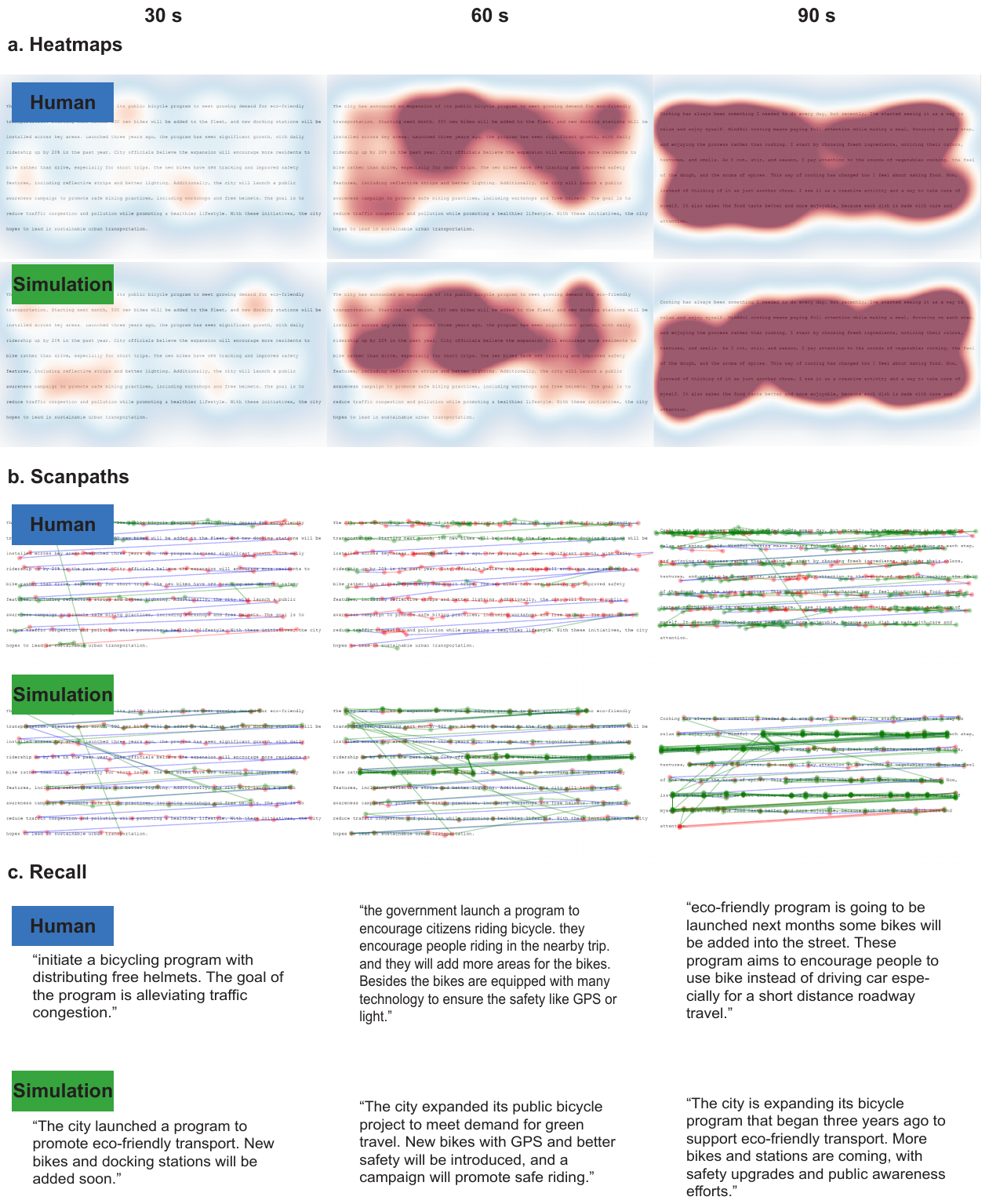} 
    \caption{
    \textbf{Reading behavior under different time constraints (30~s, 60~s, 90~s).}  
    \textbf{a. Heatmaps.} Both humans and the simulation broaden the coverage of fixations and allocate more fixation time as available reading time increases.  
    \textbf{b. Scanpaths.} Eye-movement sequences adapt to time pressure: limited time causes faster reading with more skips (blue) and fewer regressions (green), whereas abundant time allows more careful inspection and targeted regressions. Red dots and lines represent normal (first-pass) fixation points and saccades, respectively. Animated visualization examples of these scanpaths are provided in Supplementary Videos.  
    \textbf{c. Recall.} Free-recall responses reflect the same trade-off: shorter durations yield coarse recall, whereas longer durations support more complete and detailed memory.  
    Together, these patterns illustrate reading as a \emph{resource-rational} control process that flexibly adapts to time resources: when time is limited, readers optimize overall comprehension by prioritizing coverage over detail -- skipping more and regressing less; when time is abundant, they shift toward higher-accuracy comprehension by investing additional fixations and regressions.
    }
    \label{fig:scanpaths and heatmaps} 
\end{figure*}

\begin{figure*}[t]
    \captionsetup{labelfont=bf,name=Extended Data Figure}
    \centering 
    \includegraphics[width=\textwidth]{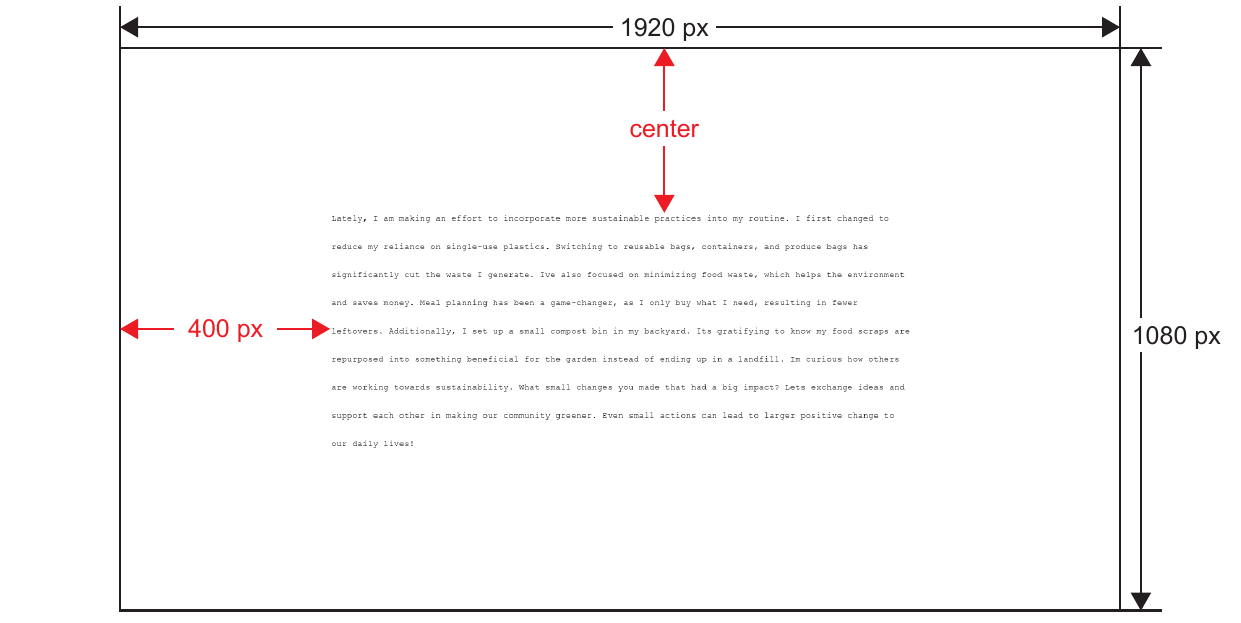} 
    \caption{
    \textbf{Stimulus presentation.}  
    Reading stimuli were displayed at the Tobii eye-tracker monitor.  
    Centering the text minimizes distortions and reduces measurement error in marginal regions, thereby improving eye-movement data quality~\cite{zermiani2024interead}.
    }
    \label{fig:extended data stimulus showcase} 
\end{figure*}

Time pressure fundamentally alters how humans read, forcing readers to trade comprehension accuracy against speed.
Prior studies have established that when readers face time constraints, they adjust reading strategies: reading speeds increase, skipping becomes more frequent, regressions are reduced, and comprehension performance declines~\cite{vibert2025impact}. However, most existing datasets either do not manipulate time directly or fail to capture the joint dynamics of gaze behavior and memory formation during English reading. To address this gap, we collected a dedicated \emph{Reading Under Time Pressure} dataset, in which 39 participants read texts (stimulus exemplified in Extended Data Fig.~\ref{fig:extended data stimulus showcase}) under three time conditions (30~s, 60~s, 90~s) while both eye movements and comprehension outcomes were simultaneously recorded. Human readers exhibited systematic adaptations across time conditions (Fig.~\ref{fig:simulation gallery}d): reading speed decreased from $192\pm56$~wpm (30~s) to $166\pm30$ wpm (90~s); skipping rate declined from $0.39\pm0.09$ to $0.31\pm0.09$; regression rate increased from $0.32\pm0.17$ to $0.59\pm0.11$, with Fig.~\ref{fig:scanpaths and heatmaps}a and b exemplify scanpaths and heatmaps under three time conditions; and comprehension performance improved with more time (multiple choice question accuracy: $0.59\pm0.23$ to $0.83\pm0.16$; free recall: $0.44\pm0.13$ to $0.52\pm0.16$, with Fig.~\ref{fig:scanpaths and heatmaps}c demonstrate free recall results). From a resource-rational perspective, these adjustments reflect a strategic trade-off of comprehension gains against time costs. When time is abundant, readers can afford to allocate more fixations to difficult texts and to regress in order to repair comprehension. As time becomes scarce, however, the value of additional fixations diminishes: forward saccades accelerate, regressions are suppressed, and attention shifts toward selectively sampling more informative regions. 

To capture these empirical adaptations, we extended the model to incorporate time constraints, framing comprehension as a resource-rational control problem where speed must be traded against comprehension performance. Unlike natural, unconstrained reading, time-pressured reading forces agents to prioritize task completion within strict temporal limits, often sacrificing local (sentence-level) or global (text-level) coherence. Our model incorporates time directly into the observation space, allowing the agent to track remaining time and adjust its behavior accordingly. Final rewards are computed based on comprehension score, which is the geometric mean of appraisals across sentences, shaped by time-sensitive factors that influence the agent’s priorities under pressure (details in Methods). This enables the agent to learn adaptive reading strategies, such as modulating skipping, regressions, and termination, based on available time. The enhanced model reproduces and explains key empirical effects (Fig.~\ref{fig:simulation gallery}d): with ample time (90~s), the agent allocated more fixations to difficult or low-confidence regions, often revisiting earlier sentences to strengthen integration. Under high time pressure (30~s), both humans and the model adopt a quantity-over-quality strategy, characterized by rapid forward saccades, minimal rereading, and selective sampling of informative regions. This shift mirrors human adaptation across conditions and reflects a strategic reweighting of expected comprehension gains against time costs: as time becomes scarce, completing the text efficiently is prioritized over further refinement of understanding.

\begin{figure*}[t]
    \captionsetup{labelfont=bf,name=Extended Data Figure}
    \centering 
    \includegraphics[width=\textwidth]{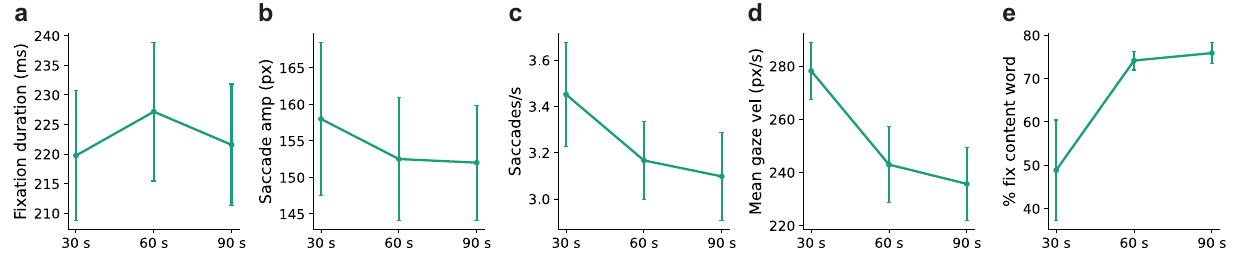} 
    \caption{
    \textbf{Simulation effects on fixation and saccade-related metrics.} 
    Each panel shows mean values $\pm$ standard deviation across three time-constraint conditions (30 s, 60 s, 90 s). Error bars represent between-episode variability within the simulated data. As time pressure decreased, fixation duration slightly lengthened, saccade amplitude and gaze velocity decreased, and the proportion of fixations landing on content words increased. These trends mirror the qualitative patterns reported in the new dataset~\cite{vibert2025impact}, even though our model was evaluated on different English stimuli and time constraints. The comparison therefore concerns the direction and relative shape of the effects, not the specific dataset or materials, demonstrating that the simulated control policies reproduce key signatures of human reading adaptation under varying time constraints.
    }
    \label{fig:extended data french corpus effects replication}
\end{figure*}

Apart from matching these empirical metrics, our model also captured the direction of effects reported in an independent eye-tracking study of reading under time constraints on a French corpus~\cite{vibert2025impact}. That study reported human adaptations consistent with our findings: as time pressure decreases, readers show longer mean fixation durations and a higher proportion of directly fixated content words, while saccade amplitude, saccade rate, and left-to-right gaze velocity decline. Our model reproduced all of these time-pressure-dependent patterns to a high level of accuracy (Extended Data Fig.~\ref{fig:extended data french corpus effects replication}).

\begin{figure*}[t]
    \centering 
    \includegraphics[width=\textwidth]{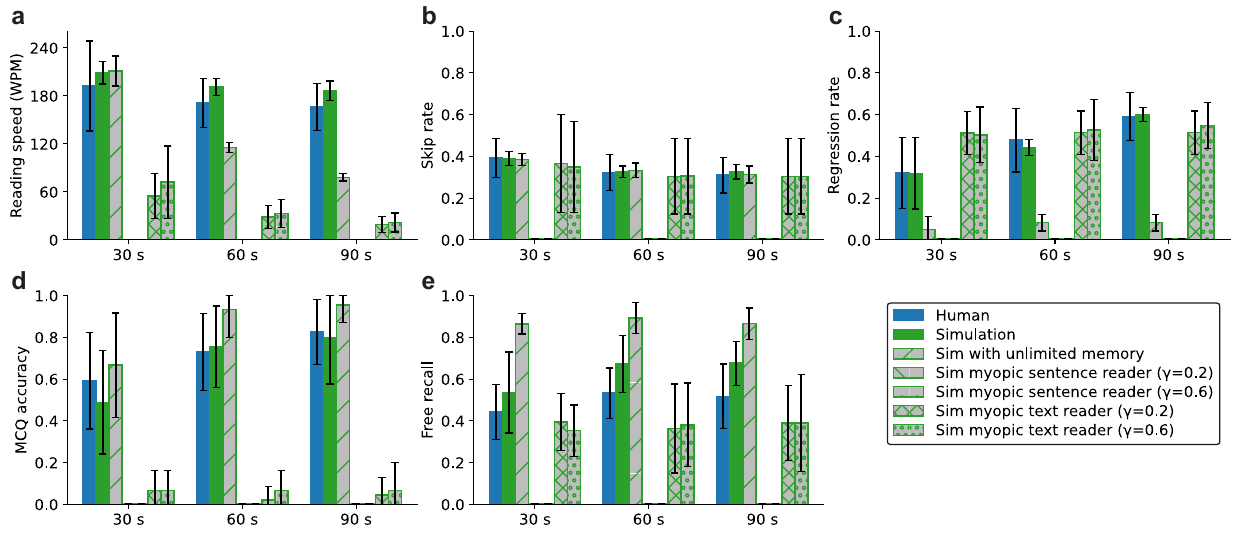} 
    \caption{
    \textbf{Effects of memory limits, hierarchical integrity, and long-term comprehension utility.}
    \textbf{a-c. Eye movements} (reading speed, skip rate, regression rate) and \textbf{d-e. comprehension} (MCQ accuracy, free recall) under three time limits (30~s, 60~s, 90~s). Human data are blue; our original model's simulation is green; modified model's simulations are grey with green hatches. Bars show means; error bars denote standard deviation and are truncated to the feasible range \([0,1]\) for bounded measures. 
    The \textit{unlimited-memory} model achieves markedly super-human MCQ and free-recall scores while showing fewer regressions and faster reading. 
    The \textit{myopic sentence reader} (\(\gamma=0.2, 0.6\)) collapses into local behavior, repeatedly fixating early words and failing to advance through the text, producing near-zero comprehension.  
    The \textit{myopic text reader} (\(\gamma=0.2\)) preserves lower-level word and sentence processing but fails to proceed reading over sentences, yielding markedly reduced comprehension; a milder myopia (\(\gamma=0.6\)) improves performance but still falls far below humans. 
    In contrast, the full agent, combining bounded memory with non-myopic optimization of future comprehension most closely matches human eye movements and comprehension. These results show that human-like reading requires realistic resource limits, sensitivity to long-term comprehension utility, and an intact hierarchical control structure.
    }
    \label{fig:comparisons} 
\end{figure*}

\subsection*{Validating the necessity of hierarchical resource rationality}

To further validate the memory-construction module design and necessity of long-term comprehension utility optimization, we compared the full hierarchical model with five baseline models (Fig.~\ref{fig:comparisons}). The first baseline, the unlimited-memory agent, stores all parsed propositions in long-term memory without strategic selection. It achieves superhuman performance: higher MCQ and free-recall scores while requiring fewer regressions and showing faster reading, confirming that human-like capacity limits are essential for generating realistic reading behavior. The next two baselines are myopic sentence readers ($\gamma=0.2,\ 0.6$), which plan only locally within each sentence, failing to optimize overall sentence-level comprehension. These agents fail to learn forward progress, producing poor comprehension and abnormal eye-movement profiles. These results are consistent with stalling at local states instead of pursuing a longer-term reading goal. The final two baselines are myopic text readers ($\gamma=0.2,\ 0.6$), which integrate information across sentences but still underperform because they lack text reader's efficient high-level planning: these agents read only the initial portions of the text and fail to decide where to read next that maximize overall text comprehension. 
Across all time conditions, only the full hierarchical model: combining bounded memory capacity and long-term value estimation, best matches human eye-movement and comprehension patterns across all time conditions. These findings reinforce that realistic reading behavior emerges from hierarchical resource-rational control: limited memory and attentional resources, together with sensitivity to long-term sequential consequences, jointly optimize comprehension utility while minimizing effort and time costs.

\section*{Discussion}

This paper establishes that human reading behavior arises from a \emph{general resource-rational control mechanism} -- one that coordinates perception, memory, and sequential action under uncertainty.
While deep reinforcement learning has been advanced as a powerful method for instantiating computational theories of the human mind~\cite{dayan2008decision,bai2024heads,shi2024crtypist,shi2025chartist,lingler2024supporting,oulasvirta2022computational}, what has been missing is an account of how hierarchical processing emerges in knowledge-intensive tasks involving eye movements.
Readers allocate fixations through strategic, sequential decisions that maximize comprehension while minimizing effort and time, under the uncertainties imposed by limited cognitive resources. 
Our model represents this principle as a hierarchical POMDP solved with deep reinforcement learning, enabling policies to emerge under realistic perceptual and memory uncertainty that faithfully capture human-like resource limits. This resource-rational hierarchical architecture bridges a long-standing divide among theories of memory, perception, and control, not only in reading research, but across behavioral and cognitive sciences. It proposes that high-level comprehension processes and lower-level processes of motor control can occur from a unified process well accounted as optimization. 
Unlike earlier models~\cite{reichle2003ez,salvucci2001integrated} that relied on handcrafted heuristics or treated gaze behavior as reflexive, our approach treats reading as an adaptive, goal-directed process of active information sampling. As a result, the model reproduces canonical behavioral phenomena across scales -- from word to sentence to text -- including effects of word length, frequency, and predictability, as well as skips, regressions, and adaptive strategies under time pressure. More importantly, it provides direct computational validation that explaining reading mechanisms through the lens of resource rationality offers a unified and general account of human reading behavior. The same trade-off between immediate and delayed informational gain appears broadly in human decision-making, suggesting that reading reveals a general cognitive strategy for allocating effort and information under uncertainty, rather than a task-specific mechanism.

resource rationality explains the remarkable adaptability observed in human reading. 
When we relaxed resource limits in simulation by granting the agent unlimited memory or full visual access to text, strategic eye-movement control disappeared. The agent no longer needed to decide where to look or when to move; comprehension could be achieved trivially without selective sampling. Conversely, when resources were bounded, the agent adapted by allocating fixations to maximize long-term comprehension utility while minimizing immediate effort. This trade-off was formalized through value estimation across future steps: when decision-making was biased toward short-term outcomes (\(\gamma=0.2,\,0.6\)), the agent became myopic and locally optimal, fixating repeatedly within a sentence without anticipating how present sampling supports later comprehension. These behavioral shifts mirror human tendencies to balance immediate versus delayed informational gain, reinforcing that reading is a goal-directed optimization process. The hierarchical architecture of our model was crucial for enabling such adaptive control. Human readers naturally decompose the complex task of understanding text into nested subtasks~\cite{reichle2003ez,rayner1998eye} -- selecting regions, words, and letters -- each operating over distinct temporal and informational scales. Such hierarchical decomposition is a signature of many complex skills~\cite{botvinick2008hierarchical}: from language comprehension to motor planning, suggesting the mechanisms uncovered here extend beyond reading. When the hierarchical structure was disrupted (e.g., by removing or collapsing any of the three levels), the agent failed to learn effective policies: 
long-term comprehension goals alone were too sparse and insufficient to guide moment-to-moment perceptual decisions without intermediate control levels.
The need for intermediate controllers illustrates how hierarchical control provides both efficiency and learnability under bounded resources. This mechanism parallels neural evidence that reading engages fast hierarchical processing of orthographic and semantic information across distributed cortical regions~\cite{wang2025fast,pallier2011cortical}.

Our approach yields clear, testable predictions about how reading behavior should vary across individuals, populations, and contexts as a function of underlying resource constraints and valuation. Differences between native and non-native readers should primarily reflect weaker lexical and predictive knowledge, leading to reduced skipping and slower integration even for frequent or predictable words~\cite{rayner2009eye}, whereas dyslexic reading should instead reflect increased perceptual and decoding uncertainty, producing longer fixations and more regressions even when predictability is high~\cite{de2002reading}. Beyond group-level effects, stable individual differences in fixation durations, skipping, and rereading should be explainable by continuous variation within a shared control architecture rather than distinct strategies. Situational factors such as time pressure~\cite{vibert2025impact} or multitasking~\cite{bai2024heads,lingler2024supporting} should modulate reading behavior by reweighting the relative value of comprehension versus efficiency, yielding predictable person-by-situation interactions. These predictions can be evaluated by fitting reader-specific constraints within a computational model~\cite{kangasraasio2019parameter}, offering a unified approach to studying proficiency, impairment, and adaptation in reading.

Our computational basis offers a plausible blueprint for complex intelligent behavior beyond reading. Many everyday activities, such as writing, sports, and skill learning, all require the coordination of high-level reasoning and decision-making with low-level perceptual-motor execution. Hierarchical control supported by multilayer memory systems provides a computational solution to such multi-level interplay. By adjusting these memory systems and applying realistic perceptual-motor constraints, the same resource-rational principles can explain a wide range of intelligent activities, offering cognitive science with a unified computational basis for understanding human behaviors across these tasks. Deep reinforcement learning further implements and enhances these principles by enabling generalizable control policies in stochastic and uncertain environments. Earlier models relied on handcrafted rules that were brittle to context or agent changes, whereas amortized training using deep reinforcement learning across varied conditions yields adaptive policies that capture human-like variability without manual tuning. This flexibility opens new opportunities for systematic evaluation of cognitive and behavioral theories in fields such as human-computer interaction, education, and psychology. For artificial intelligence, the model demonstrates how the learning of human-like behavior can emerge from decision-making rather than imitation of massive datasets. Grounding policy learning in hierarchical POMDPs integrated with cognitive theories provides a top-down route for building interpretable agents capable of reasoning, perceiving, and acting with human-like efficiency and flexibility.


\printbibliography[title={References},filter=onlymain]

\begin{refsegment} 
\section*{Methods}
\label{sec:methods}

\subsection*{Datasets}
We validated our model across five datasets spanning different levels of reading: 

\begin{itemize}
    \item Kliegl et al.’s dataset~\cite{kliegl2004length} for word recognition,
    \item ZuCo 1.0~\cite{hollenstein2018zuco} for sentence reading,
    \item McNamara et al.’s text comprehension dataset~\cite{mcnamara1996good} for text-level understanding,
    \item Vibert et al.’s French time-constrained reading dataset~\cite{vibert2025impact} for independent validation of time-pressure effects on eye movements,
    \item and a new dataset we collected for evaluating reading under time pressure.
\end{itemize}

\subsubsection*{Reading under time pressure dataset (English)}
To evaluate whether our model could reproduce human eye movements, memory and comprehension patterns under varying time constraints, we required an English reading dataset that extended beyond isolated sentences and included both eye-tracking and comprehension measures with explicit time manipulation. As no existing corpus met these requirements, we conducted a dedicated behavioral study that systematically varied available reading time (30~s, 60~s, and 90~s) while recording both gaze behavior and comprehension performance. Detailed descriptions of the datasets are provided in the Supplementary Information (Sec.~1.1.1 to 1.1.4).

We recruited thirty-nine adults and analyzed thirty-two valid participants (seven excluded for poor tracking or metric outliers), in a within-subject design manipulating available reading time (30~s, 60~s, 90~s). On each trial, participants silently read a short text on a Tobii Pro Spectrum eye tracker's display~\cite{tobiiSpectrum}, after which they completed a 20~s arithmetic distractor followed by a comprehension test (free recall $+$ 5 Multiple Choice Questions). Texts were generated to minimize prior-knowledge confounds; comprehension items were sanity-checked to avoid non-content cues. Eye movements were recorded at high frequency and parsed with Tobii’s I-VT filter~\cite{tobiiIVT}; we then applied a post hoc line-alignment correction~\cite{al2025combining} (time-warping $+$ chain filters) to improve fixation-to-line mapping. From these data we computed fixation-, saccade-, and comprehension-based metrics used to evaluate the model under the same time constraints. Full details on data collection and metric definitions could be found in the Supplementary Information (Secs.~1.1.4 and 1.2), and all materials and prompts are provided for reproducibility.

\subsection*{Model}

The model is designed to investigate three core assumptions about human reading: (i) cognitive resources for perception, memory, and time are limited; (ii) reading unfolds under uncertainty about word identity, sentence meaning, and text coherence; and (iii) readers act to optimize comprehension utility subject to these constraints. Crucially, the model does not assume fixed heuristics, hand-crafted decision rules, or task-specific thresholds for skipping, regression, or rereading. Instead, all eye-movement and rereading behaviors emerge from optimizing expected utility under partial observability and resource costs.

Specifically, human reading was modeled as a three–level resource-rational controller comprising a \wordrecognizermodel, \sentencereadingmodel, and \textreadingmodel. Each level implements a reinforcement learning (RL) policy that selects actions to maximize expected utility under cognitive and temporal costs. The controllers operate at distinct temporal scales: within-word fixations, sentence-level eye-movement control, and text-level rereading decisions, which are composed using a call-and-return execution structure.

Each controller $\ell \in \{\mathrm{W},\mathrm{S},\mathrm{T}\}$ is formalized as a Partially Observable Markov Decision Process (POMDP), specified by a tuple
\(
\langle S_\ell, A_\ell, T_\ell, O_\ell, r_\ell, \gamma_\ell\rangle ,
\)
where $S_\ell$ denotes the state representation, $A_\ell$ the action space, $T_\ell$ the transition dynamics, $O_\ell$ the observation model, $r_\ell$ the level-specific reward function, and $\gamma_\ell$ the discount factor (default $0.99$). Belief states for partially observable levels are updated using Bayesian update and memory construction.
Higher-level actions (sentence- and text-level decisions) unfold over variable durations determined by the lower-level controller and are evaluated using a hierarchical value function. This structure allows each controller to optimize its own resource-rational objective while contributing to overall reading performance.
Detailed derivations and modeling assumptions are provided in Supplementary Information Sec.~1.3, and full implementation details are available in codes.

\subsubsection*{\wordrecognizermodel}\label{method_subsubsec3}
At the lowest time scale, the agent performs a \emph{word-recognition} task. Given a word on the display, the agent selects fixation locations within the word to acquire visual information and update its belief over candidate lexical entries. Word recognition is formulated as a resource-rational control problem in which fixation actions are chosen to maximize expected lexical identification accuracy while minimizing oculomotor and temporal costs. We formalize this process as a POMDP, in which the agent integrates noisy letter observations across successive fixations to reduce uncertainty in its lexical memory.
\begin{itemize}
    \item \textbf{State \(S\).} $s=(x, I, len, w^*)$, where $x$ is current fixation letter index; $I$ is the letter information sampled so far. $len$ is the word length; $w^*$ is the ground-truth word identity. $w^*$ is the ground-truth word (not directly observable to the agent).
    \item \textbf{Action \(A\).} $a=(x^\prime, \beta)$, where $x^\prime$ is next fixation position (as letter index); $\beta\in\{\textsc{continue},\,\textsc{stop}\}$. Upon choosing \(\textsc{stop}\), the agent selects the best activated word from lexical memory according to its current belief.
    \item \textbf{Transition \(T\).}  The environment is static, only the belief $b(w)=P(w\mid I)$ is updated. Beliefs follow $b^\prime(w)\propto P(I^\prime\mid w)b(w)$. The initial prior $P(w)$ was proportional to corpus frequency (SUBTLEX-US~\cite{brysbaert2012subtlex-us}) or contextual predictability (GPT-2~\cite{radford2019language}). The likelihood $P(I\mid w)$ is computed with an ideal observer model (Supplementary Information Sec.~1.4.1). To maintain a bounded lexical memory, only the top 5 candidates are retained after each update.
    \item \textbf{Observation \(O\).} $o=(x^\prime,I_{x^\prime},len, b^\prime)$, where $x^\prime$ is the new fixation position, $i_{x^\prime}$ are letters revealed at $x^\prime$, and $b^\prime$ is the updated belief.
    \item \textbf{Reward \(r\).} $U+c(t)$, where $U$ was the final-step reward of correct (+100) or incorrect recognition (-100), and $c(t)$ was the step-wise reward (-0.1) denotes the costs of eye movements.
\end{itemize}
Unlike heuristic word-recognition models that map lexical features directly to fixation durations or stopping rules, the present formulation derives fixation timing and termination decisions from uncertainty reduction under cost.

\paragraph{Fixation and gaze durations}
The simulator produces fixation positions, which we convert into durations to enable comparison with human gaze data. Fixation time is modeled as a function of the lexical information gained during the fixation: greater reductions in lexical entropy \(\Delta H\) correspond to longer processing times, modulated by a tunable parameter \(\kappa\)~\cite{wilcox2023testing,smith2013effect} and base fixation duration \(d_0\)~\cite{rayner1998eye}. The mean lexical duration is
\[
\mu_{\text{lex}} = d_0 + \kappa \,\Delta H,
\]
clipped to the empirically observed range for single fixations of 200 to 250 ms~\cite{rayner1998eye}. Observed durations are drawn from a Gamma distribution to reproduce the characteristic right-skew of human fixation times~\cite{staub2010distributional}. 
First-pass gaze duration for a word is computed as the sum of its first-pass fixation durations~\cite{rayner1998eye}. Saccades are assigned a fixed duration of 25~ms, following standard reading models~\cite{reichle2003ez}, and are excluded from gaze duration but included in total reading time under time-pressure evaluations.
To account for non-fixation intervals (e.g., blinks, tracking losses), we inflate each fixation duration by a constant overhead factor \(\rho\)~\cite{rayner1996mindless} estimated from human data. Total time for a word is therefore
\[
D_{\text{word}} = 
\sum_{t\in\text{first-pass}}
\bigl(
d_{\text{lex}}^{\,t} (1 + \rho)
+ d_{\text{saccade}}
\bigr).
\]
Full derivations and details are provided in Supplementary Information (Sec.~1.4.1).

\subsubsection*{\sentencereadingmodel}\label{method_subsubsec1}
At the medium time scale, the agent performs a \emph{sentence-reading} control task. Given a sentence on the display, the agent selects eye-movement actions (fixate, skip, or regress) to integrate word meanings in short-term memory and form a coherent sentence interpretation~\cite{lewis2005activation,kuperberg2016we}. Sentence reading is formulated as a resource-rational control problem in which eye-movement policies are chosen to maximize expected sentence-level comprehension while minimizing time and oculomotor costs. The task is formalized as a POMDP in which the agent updates a belief over the upcoming word based on contextual expectations and parafoveal preview.
\begin{itemize}
    \item \textbf{State \(S\).}
        \(s=(j,\,\mathbf{c},\,\psi_{1:j},\,b_{\mathrm{next}})\), where
        \(j\) is the index of the currently fixated word,
        \(\mathbf{c}\) is the short-term context buffer,
        \(\psi_{1:j}\) are appraisals of the words read so far, and
        \(b_{\mathrm{next}}\) is the belief over the identity of the next word.
    \item \textbf{Action \(A\).}
        \(a=(m,\beta)\) with \(m\in\{\textsc{regress},\textsc{next},\textsc{skip}\}\) and
        \(\beta\in\{\textsc{continue},\textsc{stop}\}\).
    \item \textbf{Transition \(T\).}
        Eye-movement transitions are deterministic. The belief over the next word is updated by combining parafoveal letter information with the contextual prediction from a pretrained language model (LM):
        \[
          b^\prime_{\mathrm{next}}(w)\propto
          P(w\mid\text{letters})\,
          P_{\mathrm{LM}}(w\mid\mathbf{c}) ,
        \]
        followed by normalization. Details appear in Supplementary Information (Sec.~1.4.2).
    \item \textbf{Observation \(O\).}
        \(o=(j,\,b^\prime_{\mathrm{next}},\,\psi_j,\,u_t)\), where
        \(\psi_j\) is the appraisal of the current word and
        \(u_t\) is the running comprehension score.
    \item \textbf{Reward \(r\).}  
        Each eye-movement step incurs an oculomotor cost of \(-0.1\), scaled by a regression penalty \(w_{\mathrm{reg}}\) for \textsc{regress} actions. When the agent chooses \textsc{stop}, a terminal reward is issued: a positive reward proportional to the final sentence-level comprehension score if the sentence has been fully read, and a large negative reward otherwise. Full details are provided in Supplementary Information (Sec.~1.4.2).
\end{itemize}
Importantly, skipping and regression decisions are not triggered by lexical thresholds or structural rules, but arise from belief-guided evaluation of expected comprehension gain relative to action costs.

\subsubsection*{\textreadingmodel}\label{method_subsubsec2}
At the text level, the agent maintains a capacity-limited, noisy representation of text meaning from the imperfect memory rather than a veridical record of all previously read content.
The agent performs a \emph{text-reading} control task. Given a multi-sentence passage, the agent selects which sentence to read or revisit to construct a coherent representation of the text. Text reading is formulated as a resource-rational control problem in which sentence-selection policies are chosen to maximize expected text-level comprehension while minimizing time, memory, and re-reading costs. Because the agent does not have direct access to the true underlying text state and observes only its own memory appraisals of previously read sentences, the task is formalized as a POMDP. At each step, the agent selects a sentence to read or revisit to improve its belief about the evolving text representation.

\begin{itemize}
    \item \textbf{State \(S\).}
        \(s = (k,\,\mathbf{d},\,b_{\text{text}})\), where
        \(k\) is the index of the current sentence,
        \(\mathbf{d}\) is the static document representation, and
        \(b_{\text{text}}\) is the belief distribution over sentence-level appraisals accumulated so far.
    \item \textbf{Action \(A\).}
        \(a = (m,\beta)\), where
        \(m \in \{\textsc{next},\textsc{regress}\}\) and 
        \(\beta \in \{\textsc{continue},\textsc{stop}\}\).
        For \textsc{regress}, the agent specifies the sentence index to revisit.
    \item \textbf{Transition \(T\).}
        The text is static and eye movements land deterministically on the target sentence.
        After reading a sentence \(j\), its appraisal \(\varphi_j\) is updated using a coherence score derived from a pretrained language model, normalized to \([0,1]\). Details of the appraisal computation, integration mechanism, and normalization appear in Supplementary Information (Sec.~1.4.3).
    \item \textbf{Observation \(O\).}
        \(o = (k,\,b'_{\text{text}},\,\varphi_k,\,u_t)\), where
        \(b'_{\text{text}}\) is the updated belief over sentence appraisals,
        \(\varphi_k\) is the appraisal of the currently visited sentence, and
        \(u_t\) is the running comprehension score.
    \item \textbf{Reward \(r\).}
        Each reading step incurs a small oculomotor cost. When the agent chooses \textsc{stop}, a terminal reward is issued: a positive comprehension-based score if the entire text has been read, and a large negative penalty otherwise. The terminal score depends on the sentence-level appraisals accumulated during reading, with lower-appraisal sentences exerting disproportionately greater influence; full mathematical definitions and derivations are provided in Supplementary Information (Sec.~1.4.3).
\end{itemize}

\paragraph{Memory system}
At the text level, the agent maintains a coarse representation of text meaning, which is updated after each sentence. The sentence is first mapped to a set of conceptual units, and a small subset is selected as gist elements based on their relevance and coherence with the evolving context. These selected elements update the agent's long-term memory representation of the text, with repeated encounters strengthening their activation. Full details of the parsing procedure, schema-based selection mechanism, and long-term memory update rules are provided in Supplementary Information (Sec.~1.4.3).

\subsubsection*{Reading under time pressure}
To model the speed–accuracy trade-off imposed by time constraints, we retain the same hierarchical POMDP structure and introduce a small set of advancement at the sentence and text levels. These modifications control how time pressure alters the agent’s available steps, state representation, and reward structure.

At the sentence level, the agent receives a time-derived budget of reading steps as state and observation signals, indicating the remaining allocation and whether the agent has exceeded this budget. Once the budget is exceeded, reading continues but incurs a small overtime penalty at each step. Skipping a word leads to a controlled degradation of its integration value, reflecting the information loss observed in human skipping behavior. Early termination before all words are processed yields an additional penalty; otherwise, the usual sentence-comprehension reward is used.
At the text level, both the state and observation are augmented with the global remaining time. Episodes terminate automatically when the available time expires or the agent issues the stop action. The terminal reward combines the text-level comprehension score with a progress-dependent bonus that encourages the agent to balance depth of processing with coverage under different time constraints. Details are shown in Supplementary Information (Sec.~1.4.4).

\subsection*{Training and simulation}
\paragraph{Training}
Each controller—\wordrecognizermodel, \sentencereadingmodel, and \textreadingmodel, was trained independently in a reinforcement learning environment. Policies were parameterized by neural networks and optimized through trial-and-error interaction with their respective POMDP environments implemented in \textsc{gymnasium}~\cite{towers2024gymnasium} and trained using \textsc{Stable-Baselines3}~\cite{stable-baselines3}. This hierarchical training scheme avoids sparse-reward pathologies inherent in flat RL by assigning each level a local objective with immediate feedback. Training stimuli consisted of synthetic or corpus-derived text materials. Critically, no human behavioral data (e.g., gaze patterns or comprehension scores) were ever used during learning; all human data were reserved exclusively for post-hoc evaluation.

\paragraph{Simulation and validation}
After training, each controller was combined hierarchically to generate full reading simulations. Human behavioral datasets were used only for evaluation, enabling direct comparison between the model’s simulated eye-movement patterns and comprehension outcomes and those observed in readers. Validation followed the structure of the hierarchy: word-level simulations were assessed using gaze-duration effects, sentence-level simulations were evaluated through skip and regression behavior, and text-level simulations were assessed via recall and comprehension scores. Under time-pressure conditions, model predictions for both gaze dynamics and comprehension accuracy were compared against human adaptations to varying deadlines.
Details of training and simulation could be found in Supplementary Information (Sec.~1.5).

\subsection*{Parameters and fitting}

Most model parameters $\boldsymbol{\phi}$ were either anchored to established empirical findings or treated as tunable values optimized against behavioral benchmarks. Parameter definitions, literature justifications, ranges, and final values are provided in Supplementary Information (Sec.~1.6) and summarized in Extended Data Table~\ref{tab:tunable-params}.

Model parameters $\boldsymbol{\phi}$ were optimized to minimize a composite discrepancy between simulated and human data using a simulation-based, likelihood-free fitting procedure~\cite{lieder2020resource} (details see Supplementary Information Sec.~1.6). The discrepancy combined curve-level losses for conditional bin means (variance–aware reduced $\chi^2$), cell-mean losses when only few conditions were available (sum of squared errors), and distributional losses when full conditional distributions were available (Jensen-Shannon divergence~\cite{lin2002divergence}). Each metric subset was associated with its corresponding parameter(s): gaze-duration curves for $\kappa$, skipping/regression patterns for $w_{\mathrm{reg}}$, recall distributions for $\tau_{\mathrm{H}}$ and $\tau_{\mathrm{L}}$, and time-pressure adaptation metrics for $\rho,,w_{\mathrm{TP}},,w_{\mathrm{IL}},,w_{\mathrm{RP}}$. Fitting proceeded in stages: single-parameter subsets ($\kappa$, $w_{\mathrm{reg}}$, $\tau_{\mathrm{H/L}}$) were tuned via exhaustive grid search over specified ranges, while the multi-parameter time-pressure subset ($\rho,,w_{\mathrm{TP}},,w_{\mathrm{IL}},,w_{\mathrm{RP}}$) was optimized using Bayesian optimization~\cite{shahriari2015taking,snoek2012practical} to efficiently explore the joint space with limited simulation–evaluation cycles. The staged procedure produced a final parameter set $\boldsymbol{\phi}^*$ minimizing the composite objective and aligning simulated behavior with human benchmarks; $\boldsymbol{\phi}^*$ was shared across all analyses. All fitted parameters are theoretically grounded and optimized against aggregated human metrics, representing average rather than individual reading behavior. Details of discrepancy function and tuning procedures could be found in Supplementary Information~Sec.~1.6.3.

\printbibliography[heading=subbibliography, title={Methods references}, filter=onlymethods]
\end{refsegment}

\section*{Acknowledgments}
We thank colleagues: Prof.~Andrew Howes, Prof.~Jussi Jokinen, Prof.~Danqing Shi, Prof.~David Hsu, Dr.~Thomas Langerak, Dr.~Nuwan Janaka, Dr.~Suyog H. Chandramouli, and Alexander Lingler for helpful comments on earlier drafts. 
Y.~B., A.~O., and X.~J. were supported by the European Research Council (ERC; Grant No.~101141916) and the Research Council of Finland (Grant Nos.~328400, 345604, 341763, and 357578). S.~Z. was supported by the City University of Hong Kong (Grant No.~9610677). In addition, Y.~B. was supported by the National University of Singapore Research Scholarship and the ORIA programme, and X.~J. was supported by a scholarship from the Hong Kong University of Science and Technology (Guangzhou).

\section*{Author contributions}
A.~O. and S.~Z. supervised the project. Y.~B., A.~O., and S.~Z. designed the data collection study and developed the modeling framework. Y.~B. and A.~O. performed modeling, coding, simulations, parameter inference, data analyses and visualizations. Y.~B. and A.~O. led writing the manuscript. X.~J. collected the data. All authors revised and approved the final manuscript.

\section*{Competing interests}
The authors declare no competing interests.

\section*{Data availability}
The data supporting the findings of this study are available at the Open Science Framework (OSF) under the accession https://osf.io/q2dm6/ (DOI: 10.17605/OSF.IO/Q2DM6). Additional materials are available from the corresponding author upon reasonable request.

\section*{Code availability}
The code used to implement the model and reproduce the analyses will be made available at 
\url{https://github.com/BaiYunpeng1949/reading-modeling}. 
A permanent Zenodo DOI will be generated and provided upon publication. 
The code will be released under an open-source license.

\section*{Additional Information}
\subsection*{Supplementary information}
Supplementary Information is available in the submitted PDF file and will be made available online upon publication.
\subsection*{Correspondence and requests for materials} should be addressed to Antti Oulasvirta.

\newpage

\begin{table}[t]
    \captionsetup{labelfont=bf,name=Extended Data Table} 
    \centering
    \caption{Assumptions and tunable parameters in the reading model. L stands for Literature, O stands for Optimization.}
    \label{tab:tunable-params}
    \begin{tabular}{@{}llp{5.2cm}ccp{1.4cm}@{}}
    \toprule
    \textbf{Type} & \textbf{Name} & \textbf{Explanation} & \textbf{Range} & \textbf{Value} & \textbf{Method} \\
    \midrule
    \multirow{2}{*}{Visual perception} 
    & $n_{fov}$ & Number of high-acuity letters visible in foveal vision & $[7,9]$ letters & $8$ letters & L \\
    \cmidrule(lr){2-6}
    & $n_{parafov}$ & Number of previewable letters in the next word (parafoveal vision) & $\leq3$ letters & $2$ letters & L\\
    \midrule
    \multirow{3}{*}{Eye movement} 
    & $d_{saccade}$ & Duration for executing a saccade & $[20,50]$ ms & $25$ ms & L \\
    \cmidrule(lr){2-6}
    & $\kappa$ & Processing time cost per bit of lexical information gain (entropy change) & $(2,4)$~ms/bit & 2.50~ms/bit & O \\
    \cmidrule(lr){2-6}
    & $\rho$ & Non-fixation overhead fraction & [0.1,0.3] & 0.29 & O \\
    \midrule
    \multirow{2}{*}{Memory} 
    & $C_{STM}$ & Capacity of the short-term memory buffer & $[5,9]$ & 5 & L \\
    \cmidrule(lr){2-6}
    & $\tau_{\mathrm{LTM}}$ & Reader's proposition activation threshold for integrating into the LTM & $[0,1]$ & 0.32 & O \\
    \cmidrule(lr){2-6}
    & $\tau_{\mathrm{H}}$ & High prior knowledge readers' proposition activation threshold for merging into LTM & $[0,1]$ & 0.838 & O \\
    \cmidrule(lr){2-6}
    & $\tau_{\mathrm{L}}$ & Low prior knowledge readers' activation threshold & $[0,1]$ & 0.844 & O \\
    \midrule
    Information loss & $w_{IL}$ & Penalty weight for information loss due to skipped words & $[0,1]$ & 0.70 & O \\
    \midrule
    \multirow{2}{*}{Preferences} 
    & $w_{reg}$ & Weighting the cost of regressions & $[0,1]$ & 0.80 & O \\
    \cmidrule(lr){2-6}
    & $w_{RP}$ & Weighting for forward reading progress over regressions & $[0,3]$ & 1.30 & O \\
    \bottomrule
    \end{tabular}
\end{table}

\end{document}